\documentclass[10pt, conference, letterpaper]{IEEEtran}
\IEEEoverridecommandlockouts
\usepackage{cite}
\usepackage{amsmath,amssymb,amsfonts}
\usepackage{url}
\usepackage{algorithmic}
\usepackage{algorithm} 
\usepackage[french,boxed,vlined,linesnumbered,inoutnumbered,rightnl,algo2e]{algorithm2e}
\usepackage{graphicx}
\usepackage{textcomp}
\usepackage{xcolor}
\usepackage{comment}
\usepackage{booktabs}
\usepackage{subcaption}
\usepackage{multirow}
\usepackage{mathrsfs}
\usepackage{threeparttable}
\usepackage{booktabs}
\newcommand*{\note}[1]{\textcolor{red}{#1}}

\newcommand{\algname}{FedHC\xspace}

\def\BibTeX{{\rm B\kern-.05em{\sc i\kern-.025em b}\kern-.08em
    T\kern-.1667em\lower.7ex\hbox{E}\kern-.125emX}}

\begin{document}

%\title{An Adaptive Quantity Approach for Minimizing Cost and Delay in Satellite Clustered Federated Learning}

%\title{An Adaptive Quantization Approach for Cost and Latency Minimization in Satellite Clustered Federated Learning Framework \note{[TENTATIVE]}\\

%\title{An Adaptive Quantization Approach for Multi Objective Optimization in Satellite Clustered Federated Learning Framework}

%\title{Satellite Clustered Federated Learning Framework with Adaptive Quantization \note{[TENTATIVE]}\\

\title{\algname: A Hierarchical Clustered Federated Learning Framework for Satellite Networks \\

}

%\author{Anonymous Authors}

%\begin{comment}
\author{\IEEEauthorblockN{Zhuocheng Liu$^{1}$, Zhishu Shen$^{1}$\textbf{\IEEEauthorrefmark{2}}\thanks{\IEEEauthorrefmark{2} Corresponding author (z\_shen@ieee.org).}, Pan Zhou$^{1}$, Qiushi Zheng$^{2}$, and Jiong Jin$^{2}$}

\IEEEauthorblockA{\textsuperscript{$^{1}$}School of Computer Science and Artificial Intelligence, Wuhan University of Technology, China\\
}

\IEEEauthorblockA{\textsuperscript{$^{2}$}School of Science, Computing and Engineering Technologies, Swinburne University of Technology, Australia\\
}
%\IEEEauthorblockA{E-mail: \{\note{xxx}, \note{xxx}, yjl\}@whut.edu.cn, z\_shen@ieee.org, xxx@xxxx.com, jiongjin@swin.edu.au}
}
%\end{comment}

\maketitle

\begin{abstract}
%With the proliferation of data-driven services, the volume of data that needs to be processed by satellite networks has significantly increased. Federated learning (FL) is well-suited for handling big data processing at distributed, resource-constrained satellite environments. However, ensuring the convergence performance of FL models within dynamically distributed satellite networks is challenging while minimizing FL processing time and energy cost. To this end, we propose a hierarchical clustered federated learning framework named \algname. In this framework, a satellite-clustered parameter server (PS) selection algorithm is deployed at the cluster aggregation stage to partition nearby satellites into distinct clusters, designating the satellite cluster center client as the PS to accelerate the model aggregation. Several communicable cluster PS satellites are then selected through ground stations to aggregate global parameters to support the FL process. Moreover, a meta-learning-driven satellite re-clustering algorithm is introduced to enhance adaptability to dynamic satellite cluster changes. The extensive experiments conducted on satellite networks testbed demonstrate that \algname can significantly reduce processing time (up to 3x) and energy cost (up to 2x) against other comparative methods, while ensuring the accuracy of the model.

With the proliferation of data-driven services, the volume of data that needs to be processed by satellite networks has significantly increased. Federated learning (FL) is well-suited for big data processing in distributed, resource-constrained satellite environments. However, ensuring its convergence performance while minimizing processing time and energy consumption remains a challenge. To this end, we propose a hierarchical clustered federated learning framework, \algname. This framework employs a satellite-clustered parameter server (PS) selection algorithm at the cluster aggregation stage, grouping nearby satellites into distinct clusters and designating a cluster center as the PS to accelerate model aggregation. Several communicable cluster PS satellites are then selected through ground stations to aggregate global parameters, facilitating the FL process. Moreover, a meta-learning-driven satellite re-clustering algorithm is introduced to enhance adaptability to dynamic satellite cluster changes. The extensive experiments on satellite networks testbed demonstrate that \algname can significantly reduce processing time (up to 3x) and energy consumption (up to 2x) compared to other comparative methods while maintaining model accuracy.

%Low Earth Orbit (LEO) satellites need to process vast amounts of communication data from the ground daily. Federated Learning (FL) is well-suited for handling this data at the edge, where resources are limited. FL only requires the transmission of trained model parameters during data processing, reducing communication needs while achieving good accuracy. However, maintaining the convergence performance of FL models in a dynamically distributed satellite architecture, while minimizing communication time and energy consumption, remains a critical challenge.In this work, we propose a novel FL framework called hierarchical clustered quantitative federated learning (\algname)., to address the aforementioned issues while ensuring FL learning accuracy. \algname is built on two stages of operation: satellite cluster aggregation stage and ground station aggregation stage. We use K-means clustering algorithm to divide nearby satellites into different clusters, and specify the client of the satellite cluster center with good communication as the parameter server to accelerate the aggregation speed of the model. Aggregate global model parameters from different satellite clusters to perform FL. After the aggregation process is completed, the ground station selects a communicable cluster parameter server satellite to aggregate global parameters. Conduct model experiments on real satellite distribution data to verify the effectiveness of \algname. The results indicate that \algname can significantly reduce communication time and energy consumption with the same accuracy.
\end{abstract}


\section{Introduction}

%With the rapid development of communication technology and the proliferation of Internet of Things (IoT) devices, the computing task collected at the network edge devices has significantly increased. Distributed computing has emerged as a promising paradigm to exploit the abundant and massive computation resource of edge devices collaboratively~\cite{ChenINFOCOM23}. Among distributed computing schemes, federated learning (FL) only requires the transmission of trained model parameters during the data processing. This scheme can maintain data localization at the end-users throughout the collaborative learning process of the global model, by which enhances the security of user data. Numerous existing research have already focused on the implementation of FL in diverse wireless networks at the ground~\cite{LiaoINFOCOM23,LiTWC24}.

With the rapid development of communication technology and the proliferation of Internet of Things (IoT) devices, the computing task collected at the network edge devices has significantly increased. However, existing terrestrial wireless networks cover only a limited area of the Earth due to the geographic and economical constraints on deploying commercial mobile network infrastructure. Satellite networks serve as a viable complement to terrestrial networks, offering global coverage and seamless connectivity to support diverse computing tasks from both ground-based and space-based sources~\cite{MahboobCST24,PengISCC24}. %\textcolor{blue}{Satellites continuously collect high-resolution Earth observation images and sensor data every day, which can support a wide range of applications, such as fire detection [to be update ...] and disaster prediction [...]. Among these, post-disaster building damage assessment is considered one of the most important applications.}
Federated learning (FL) is a promising distributed learning paradigm for satellite networks, with significant potential for widespread deployment. This approach improves data processing efficiency by enabling distributed data analysis on independent satellites without requiring raw data sharing, thus preserving data privacy~\cite{ChenWC22,ShenCSUR23,FontanesiCST25}. The effectiveness of FL in supporting basic data processing tasks on real satellite computing platforms has already been demonstrated in~\cite{TangTMC24}. While FL addresses data sharing and privacy concerns, it is essential to design an FL framework tailored for satellite networks to achieve efficient collaborative model training.

%Although FL mitigates data sharing and privacy concerns, challenges remain in managing processing time and energy consumption, particularly in resource-constrained satellite networks with dynamically changing topologies.

The traditional method to conduct FL in satellite clusters typically involves two training phases. First, local training is performed on individual satellite clients. Then, collaborative model training takes place through centralized \textbf{parameter servers (PS)} on the ground. During this process, each client trains its model using its local data and transmits intermediate results, such as gradients and weights, between satellites for global model aggregation~\cite{ChenPIMRC23}. Although this FL method eliminates the need for raw data sharing to address privacy concerns, aggregating all intermediate results at a centralized PS in resource-limited satellite networks can lead to substantial communication overhead and energy consumption. Additionally, satellite network topology is constantly changing due to the high orbital speeds of satellites. Depending on the satellite type, the orbital period can range from 2 to 24 hours. While ground-based PS have abundant processing and communication resources to facilitate the FL process, communication with individual satellites is limited to specific time windows throughout the day~\cite{MathIN24}.

To address this issue, recent research has focused on implementing PS on satellites to enable distributed FL process. Most current research that considers satellites as PS typically groups satellites within the same orbit into clusters, performing FL within each group. For example, Zhai \textit{et al.} designed a distributed FL framework utilizing the Ring all-reduce algorithm to synchronize FL models and accommodate the circular topology of satellite constellations~\cite{ZhaiTMC24}. Similarly, Yan \textit{et al.} optimized the number of satellites and orbits in each cluster to minimize the convergence time of distributed FL~\cite{YanTVT24}. Distributed FL aggregates model parameters through peer-to-peer communication between clients without relying on a PS. However, these methods can lead to excessive communication between clients, resulting in inefficient use of satellite resources and increased communication costs. Cluster FL combines the advantages of both traditional and distributed FL by grouping clients into distinct clusters and performing FL independently within each cluster~\cite{LiuICC20,FedCE23}. Designing an FL solution that effectively harnesses large-scale, resource-constrainted, and fast-moving satellite networks is crucial for minimizing processing time and energy usage while ensuring model accuracy.

%Developing an FL solution that efficiently leverages the massive resource-constrained \note{vast yet resource-constrained?} and rapidly moving satellite network is crucial for minimizing processing time and energy consumption while maintaining satisfactory model accuracy.

%Razmi \textit{et al.} proposed an satellite resources  utilization scheme in a low Earth orbit (LEO) satellite cluster network~\cite{RazmiTOC24}. 
%In addition, the varying computing capabilities of edge devices require adaptive and efficient quantization techniques to ensure model performance. 
%Furthermore, a weight parameter has been incorporated to dynamically assess the necessity of quantizing the global parameters transmitted by the model, and to determine the appropriate number of quantization bits based on gradient changes.
%Here, the delay aggregation quantization (DAQ)~\cite{FartashNeurIPSs20} is employed to bypass unnecessary parameter uploads. This is achieved by estimating gradient innovation, which represent the differences between the current unquantized gradient and the previously quantized gradient.

To this end, we propose a hierarchical clustered federated learning framework, \algname, designed to optimize both processing time and energy consumption in satellite networks. The problem of clustered FL optimization with parameter server selection is NP-hard~\cite{XuIoT24}. To address this, we divide the FL process into two stages: 

\textbf{In the satellite cluster aggregation stage}, the framework first captures the location information of satellite clients and applies a clustering algorithm to group nearby satellites into distinct clusters. The satellite at the cluster center is designated as the PS to accelerate model aggregation. 
Each cluster independently aggregates local model parameters before contributing to the global FL process.
%Different satellite clusters aggregate the global model parameters for performing FL. \note{seems unnecessary}

\textbf{In the ground station aggregation stage}, the ground station selects a subset of cluster PS satellites that can communicate to aggregate global parameters. However, due to the high-speed movement of satellites, the network topology continuously changes, potentially altering cluster compositions. 
%\note{please help me rewrite this sentence to make sure the same meanings as the previous version, This dynamic environment may necessitate re-clustering, which necessitates a re-assessment of the cluster composition.} 
In such a dynamic environment, re-clustering becomes unavoidable, compelling a re-assessment of the cluster’s composition to adapt to shifting conditions. After re-clustering, mismatches between existing and newly added satellites may lead to slower convergence and increased energy consumption. To tackle this issue, we integrate model-agnostic meta-learning (MAML) after re-clustering. MAML enables the model to rapidly adapt to new cluster parameters by leveraging prior experience from different clusters. This adaptation helps maintain high performance in dynamic network environments while reducing convergence time and energy consumption.

The main contributions of this paper are as below:
\begin{itemize}
    %\item \textcolor{blue}{For disaster remote sensing data within satellite regions, we propose a self-supervised federated learning framework for efficient satellite image training. Additionally, our framework incorporates a tailored clustering strategy to reduce processing time and energy costs when deploying FL in dynamic satellite networks.}
    \item A hierarchical clustered FL framework \algname is proposed to reduce the processing time and energy consumption in dynamic satellite networks. \algname organizes the satellite FL process into two stages: satellite cluster aggregation and ground station aggregation. Initially, satellite clusters train FL models locally, and then global model updates are consolidated at the ground station, minimizing communication time.

    %a hierarchical clustered Federated Learning (FL) framework to address the challenges of processing time and energy consumption in dynamic satellite networks. This framework organizes the satellite FL procedure into two phases: 1) intra-cluster aggregation, where satellite clusters independently train localized FL models, and 2) ground station aggregation, where global model updates are consolidated at the ground station. By prioritizing local training within clusters before transmitting refined updates to the ground, the framework minimizes communication overhead and accelerates convergence in resource-constrained orbital environments

    \item A meta-learning-driven satellite re-clustering algorithm is developed to enhance adaptability to dynamic satellite cluster changes during the FL process. This algorithm enables newly joined satellites to leverage knowledge from previous cluster members, facilitating rapid adaptation to evolving network conditions and accelerating model convergence.
    
    %we integrate meta-learning into the existing federated learning framework, allowing new nodes to leverage the experience of previous ones. This enables efficient scenario adaptation and significantly accelerates model convergence.
    
    \item The effectiveness of \algname is verified by the numerical experiments. The obtained results demonstrate that \algname outperforms other comparative methods in terms of accuracy, processing time and energy consumption. 

\end{itemize}

%The remainder of this paper is organized as follows:  related work is summarized in Section~\ref{sec:relatedwork}. Section~\ref{sec:model} presents the system model and formulates the optimization problem. Section~\ref{sec:algorithm} describes our proposed clustered FL framework \algname and its associated algorithms. Section~\ref{sec:evaluation} shows the experimental results that validates the performance of \algname. Finally, conclusion with the future direction is provided in Section~\ref{sec:conclusion}.
%\input{2_relatedwork}
\section{System Model}\label{sec:model}

%如图1所示，我们考虑在数量为C的卫星集群网络中，将卫星网络划分为K个区域，每个区域选择一个LEO卫星当做参数服务器，来对集群内的卫星客户端进行联邦学习训练，地面站选择gpk个参数服务器作为它的客户端进行全局聚合

\subsection{Network model}
\begin{figure}[tb!]
\centerline{\includegraphics[width=1\linewidth, height=0.6\linewidth]{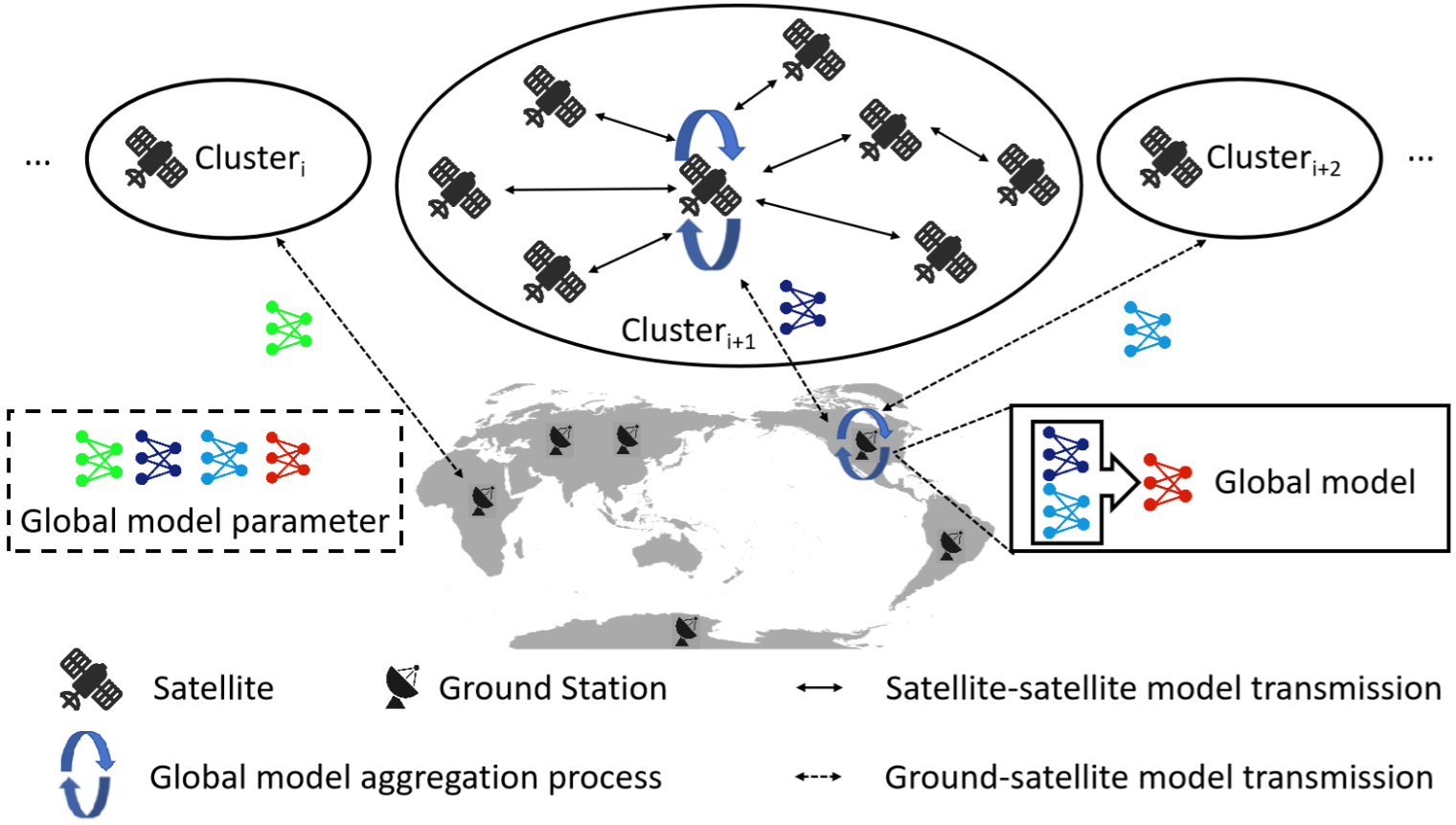}}
\caption{The network model with federated learning.}
\label{fig:model}
\end{figure}

\figurename~\ref{fig:model} illustrates the assumed satellite network model, which consists of multiple LEO satellites operating at altitudes between 500 and 2000 km. These satellites work in conjunction with ground stations to provide communication services. In this paper, we designate satellites as PS, eliminating the need to download local satellite data to the ground.
The satellite network is divided into $K$ regions (\textbf{clusters}), where each cluster contains $C^k$ satellites (\textbf{clients}). Each ellipse in \figurename~\ref{fig:model} represents an independent satellite cluster. Within each cluster, one satellite is selected as the PS to perform FL training on the satellite clients within the cluster. Once FL training is completed within the satellite clusters, each ground station $G$ selects satellite PS and aggregates their models globally. After global aggregation, the updated model parameters are transmitted back to the corresponding satellite PS, completing the FL process.

%After completing FL training within the satellite cluster, the ground station $g$ selects no more than $K$ satellite PS as its clients for global aggregation, where $p_k$ represents the satellite cluster PS to which the ground station belongs.

We assume that the ground stations operate independently without intercommunication. Each ground station has visibility of all satellites within its assigned cluster, and satellite motion follows a certain periodicity. %and we divide the network topology period into $\sigma $ time slots.
%, each with a duration of $u$. During each time slot the satellite network topology remains constant, allowing it to be treated as a quasi-static topology\cite{DuTAES17}.

\subsection{Federated learning model}
The objective of satellite FL is to train a shared learning model that enables collaboration among multiple satellite clients without transmitting local private data to a central parameter server. Specifically, the learning objective is to determine suitable model parameters $\mathbf{w}$ to predict the data for input information. We consider a satellite network comprising $C$ clients participating in FL, where each client $i \in C$ possesses a local dataset $D_i$ of size $|D_i|$. 
\begin{comment}
The local loss function of satellite client $i$ on all its data is defined as: 
\begin{equation}
F_i(x) = \frac{1}{|D_i|} \sum_{\Lambda \in D_i} f_i(w; \Lambda)
\end{equation}
where $f_i(w; \Lambda)$ is the loss function calculated by a specific sample $\Lambda$. 
\end{comment}
Therefore, the global loss function can be represented as
\begin{equation}
F(x) = \frac{1}{|D|} \sum_{i \in C} \sum_{\Lambda \in D_i} f_i(w; \Lambda)
\end{equation}
where $f_i(w; \Lambda)$ is the loss function calculated by a specific sample $\Lambda$, and $|D|=\sum_{i \in C}|D_i| $ is the total dataset size across all clients. %The objective of FL is to determine the optimal model parameters $w^* = \arg \min_w L(w)$, that minimize the global loss function. 
The objective of FL is to determine the optimal model parameters $w^*$ that minimizes the global loss function as: 
%\begin{comment}
\begin{equation}
w^* = \arg \min_w L(w)
\end{equation}
%\end{comment}

In each aggregation round $m$, the satellite PS broadcasts the global model \(w_m\) to all participating clients. Each client $i$ performs a local gradient descent to update its local parameter \(w_m^i\). During each FL aggregation round, the clients update their parameters over $\lambda$ epochs of local training. The gradient is then computed using Stochastic gradient descent (SGD) as:
\begin{equation}
    \nabla \tilde f_i(w_{m,\lambda}^{i}) = \nabla  l(w_{m,\lambda}^{i}; \Lambda)
\end{equation}

For each epoch $\lambda$, the evolution of local model parameter \(w_{m,\lambda}^{i}\) by:
\begin{equation}
w_{m,\lambda+1}^{i} = w_{m,\lambda}^{i} - \eta \nabla \tilde{f}_i(w_{m,\lambda}^{i})
\label{eq:evo}
\end{equation}
where $\eta$ represents the learning rate of model. 

\begin{comment}
and the local gradient is defined as:
\begin{equation}
 \nabla \tilde{f}_i(x_{t,\tau}^i) = \nabla L(x_{t,\tau}^i; \Lambda_{t,\tau}^i)   
\end{equation}
\end{comment}

Upon completing the local update, the client sends the weight $w_m^i$ to the FL processing server. The server subsequently aggregates these weights to create the global model parameter for the next FL round, expressed as:
\begin{equation}
w_{m+1} =  \sum_{i \in C} \frac{D_i}{D} w_m^i
\end{equation}

\subsection{Satellite cluster problem formulation}

\textbf{Processing time analysis:} We assume that at a given time $t_0$, multiple satellite clusters perform local training in parallel, with each client $i$ holding a local dataset  $D_i$. The computing power (i.e. CPU frequency) of client $i$ is denoted by $f_i$ and $Q$  represents the number of CPU cycles required to train a single data sample. Therefore, the computation time for client $i$ is given by $t_{i}^{cmp}= {D_i Q_i}/{f_i}$.
\begin{comment}
\begin{equation}
t_{i}^{cmp}= \frac{D_i Q_i}{f_i}
\end{equation}
\end{comment}

After local training, the client $i$ uploads its local models to the satellite PS for global model aggregation. The achievable transmission rate of client $i$ is expressed as:
\begin{equation}
r_i = B_i \ln \left(1 + \frac{P_0 h_i}{N_0}\right)
\end{equation}
where $N_0$ represents background noise power, $P_0$ represents transmission power, $B_i$ is transmission bandwidth, and $h_i$ is channel gain. We assume that the size of data that each client needs to upload is a constant  $\zeta$, so the communication time of client $i$ is $t_{i}^{com} = {\zeta}/{r_i}$. We further use $T^j_i$ to represent the entire training time of the $i$-th client at communication round $j$ , including computation and communication time as $T_{i}^{j} =  t_{i}^{\text{cmp}} + t_{i}^{\text{com}} $.
\begin{comment}
\begin{equation}
T_{i}^{j} = \tau_{i}^{j} ( t_{i}^{\text{cmp}} + t_{i}^{\text{com}} )
%\nonumber 
        = \tau_{i}^{j}( \frac{D_i Q_i}{f_i} + \frac{\zeta}{B_i \ln(1 + \frac{\rho_i h_i}{N_0})} )
\end{equation}
\end{comment}

In the synchronized FL system, the global model aggregation starts only when the server receives updates from all participating clients. Hence, the training time for the communication round j, indicated by $T_j$, can be calculated by $T_j = \max_{i \in C_j} \{ T_{i}^{j} \}$.
\begin{comment}
\begin{align}
T_j &= \max_{i \in C_j} \{ T_{i}^{j} \}
\end{align}
\end{comment}
It is worth noting that the value of $T_j$ is largely determined by the slowest client in $C_j$.
Beyond the aggregation time within the satellite cluster, the communication time for parameter server $p_k$ to aggregate and broadcast the global model back to its satellite cluster must also be considered. The satellite PS of the ground station is denoted as $g_{p_k}$. The satellite PS of the $i$-th cluster is $K_i^a$. Therefore, the total processing time of FL can be expressed as:
\begin{equation}
T_c = \sum_{K_i^a \in {g_{p_k}} } T_j = \sum_{K_i^a\in {g_{p_k}}} \max_{i \in C_j} \left( \max_{i \in C^k}  \left(t_{i}^{\text{cmp}} + t_{i}^{\text{com}}\right) + t^{\text{com}}_j \right)
\label{eq:time}
\end{equation}

\textbf{Energy consumption analysis}:
The energy consumption in FL can be divided into two parts: the first part is the transmission energy consumption, which arises as satellites upload their local models to the satellite parameter server.  After global aggregation at the ground station, the updated model must be broadcast back to each satellite client. Therefore, the transmission energy consumption for satellite model distribution, denoted as $E_{\text{tr}}$, is defined as:
\begin{equation}
E_{\text{tr}} = \sum_{i \in C} P_0 \frac{|w_i|}{r_i}
\end{equation}
where $|w_i|$ is the size of the global model parameters for the $i$-th client. Subsequently, the aggregation energy \(E_{\text{agg}}\) in the satellite cluster can be derived as:
\begin{equation}
E_{\text{agg}} = \sum_{i \in K} \epsilon_0 f_i t_i^{\text{cmp}}
\end{equation}
where $\epsilon_0$ is the constant coefficient determined by the hard-
ware architecture. Therefore, the total energy consumption is: %can be defined as follows:
\begin{equation}
E_c= \min ( E_{\text{tr}} + E_{\text{agg}}) 
\label{eq:energy}
\end{equation}

\begin{comment}
s.t.
\begin{subequations}

\begin{equation}
x_a \in \{0, 1\}, \ \forall a \in A,
\end{equation}
\begin{equation}
y_v \in \{0, 1\}, \ \forall v \in V,
\end{equation}

\end{subequations}

\begin{equation}
r \in \{v_{\text{l,T}} | l \in L\},
\end{equation}

\begin{equation}
y_r = 1,
\end{equation}

\begin{equation}
x_a \in \{0, 1\}, \ \forall a \in A,
\end{equation}

\begin{equation}
y_v \in \{0, 1\}, \ \forall v \in V,
\end{equation}

\begin{equation}
x_a \leq y_v, \ x_a \leq y_{v'}, \ \forall a = (v, v') \in A.
\end{equation}
\end{comment}

\textbf{Objective function}: The objective of this problem is to minimize both processing time $T_c$ (Equation~\ref{eq:time}) and energy consumption  $E_c$ (Equation~\ref{eq:energy}) generated during the FL process. Given the constraints of satellite communication and computational capacity, the objective function $F(C,K)$ can be formulated as:
\begin{align}
F(C,K) &= \min  (T_c , E_c)
\end{align}

%在处理FL的整个过程中，在满足所有数据不超过，卫星的计算容量上限和通信容量上限的前提下，我们objective function 可以被表示为

\iffalse
s.t.
\begin{subequations}
\begin{equation}
x_a \in \{0, 1\}, \ \forall a \in A,\label{eq:x_a}
\end{equation}
\begin{equation}
y_v \in \{0, 1\}, \ \forall v \in V,\label{eq:y_v}
\end{equation}
\begin{equation}
\Psi \in \{0, 1\}
\end{equation}
\begin{equation}
     0 \leq t_{i} \leq t_{\text{max}},\label{eq:t}
\end{equation}

%\begin{equation}
%     0 \leq p_{\tau_{n,k}} \leq P_{\text{max}},
%\end{equation}
\
\end{subequations}
where constraints~\ref{eq:x_a} and \ref{eq:y_v} confirm that $x_a$ and $y_v$ are binary decision variables, while $\Psi$ is a weight parameter. Constraint~\ref{eq:t} denotes that the transmission delay should not exceed the maximum allowable delay.

\fi
\section{Hierarchical Clustered 
Federated Learning } \label{sec:algorithm} 

\subsection{Overview}

\begin{figure}[tb!]
\centerline{\includegraphics[width=1\linewidth]{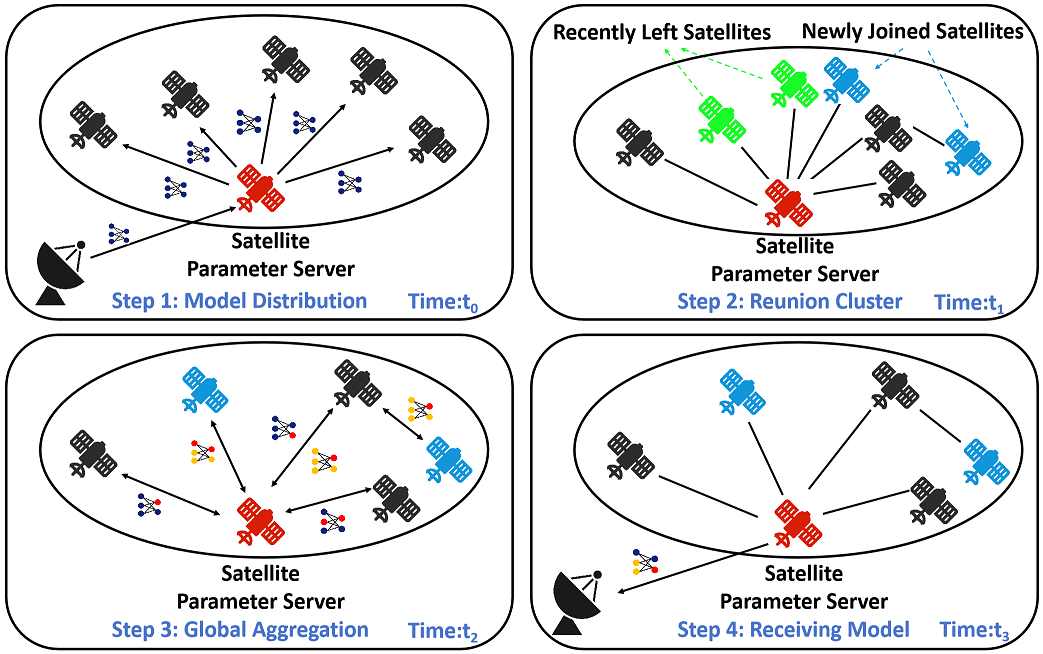}}
\caption{Overview of the proposed framework \algname.}
\label{fig:overview}
\end{figure}

\begin{algorithm}
\begin{algorithmic}[1]
\caption{\algname for satellite networks}\label{alg:decentralized_FL}
    \REQUIRE{Information of satellite networks, $K$, $C$}
    \ENSURE{Global model $w_{G}$}

\STATE Conduct satellite-clustered parameter server selection algorithm;  \label{line:1}

\STATE \COMMENT{Satellite client initialization model parameters}  \label{line:2}
\FOR {$\forall$ satellites}
     \STATE Initialize global model parameter $w_0$;
\ENDFOR  \label{line:5}

%\STATE All satellites: Initialize global model parameter $w^{(a)}_0$;

\FOR {each FL round $m \in M$}  \label{line:6}
    \STATE \COMMENT{Train local model (in-orbit computing)}
    \FOR {each satellite $i \in C$ in parallel}
        \STATE $w_{m,\lambda+1}^{i} = w_{m,\lambda}^{i} - \eta \nabla \tilde{f}_i(w_{m,\lambda}^{i})$;
    \ENDFOR \label{line:9}
    %\STATE \note{Conduct adaptive weight quantification algorithm to obtain $w{_{m}^{i}}^{\prime}$;}
    \STATE \COMMENT{Aggregate satellite cluster models}
    \FOR {each satellite $i \in C^k$ in the same cluster}  \label{line:12}
        \STATE $w_{m+1} = w_m + \sum_{i \in C^k_i} p_i w_m$;   \label{line:13}
    \STATE \COMMENT{Check if it is necessary to reassemble clusters}  \label{line:14}
        \STATE Calculate dropout rate: $d_r = \frac{C^d}{C^k}$
        \IF {$d_r > Z$}
            \STATE Re-cluster the satellites
        \ENDIF  \label{line:18}
    \ENDFOR
\ENDFOR  
\STATE \COMMENT{Aggregate global model at the ground station}  \label{line:21}
\FOR {each PS $G_{k}, k \in {K}$ in different satellite clusters}
    \STATE $w_{G} =  \sum_{k \in {K}} \frac{D_k}{D} w_m^k$;  
\ENDFOR  \label{line:23}
\RETURN $w_{G}$.
\end{algorithmic}
\end{algorithm}

\figurename~\ref{fig:overview} illustrates the flowchart of our proposed \algname. The clustering FL clustering process in \algname includes two stages: \textbf{satellite cluster aggregation stage} (Step 1-3) and \textbf{ground station aggregation stage} (Step 4). In the satellite cluster aggregation stage, a clustering algorithm is introduced to divide the satellites into distinct clusters. Within each satellite cluster $K_i^a$, the algorithm selects a satellite near the cluster center with strong communication capabilities to act as the PS. The PS is responsible for aggregating model parameters from satellites within its cluster and establishing communication with the corresponding ground stations (Step 1). During the local training process, satellites may dynamically join or leave a cluster, necessitating re-aggregation. To address this issue, MAML is introduced to adjust the initial model parameters of the newly joined satellites, allowing them to better adapt to the tasks of the new cluster (Step 2). This method accelerates the overall convergence of the satellite PS aggregation process. Following each training round, the PS combines parameters from all satellites within its cluster and distributes the updated, aggregated parameters back to them (Step 3).

%This approach facilitates the acceleration of the overall convergence of the satellite PS aggregation process. After each training round, the PS aggregates the parameters of the satellites in its cluster and distributes the \note{Do we need to add ''updated" here?} aggregated parameters to them (Step 3). 

After a specified number of training rounds in the satellite cluster, the ground station aggregation stage starts. In this stage, the ground station communicates with visible satellite clusters to aggregate their model parameters of the respective satellite clusters they are affiliated. Finally, the ground station returns the trained model parameters to the respective satellite clusters (Step 4).

As detailed in Algorithm~\ref{alg:decentralized_FL},  we introduce a satellite-clustered parameter server selection algorithm to partition the original satellite network into distinct satellite clusters based on the satellite network information. For each cluster, the algorithm selects a satellite near the cluster center with robust communication capabilities as the PS (line \ref{line:1}). Then we initialize the global model parameters $w_0$ for all satellite clients within each cluster (lines \ref{line:2}-\ref{line:5}). During each round of FL aggregation, the local satellite client first performs local training to update the global model parameters $w_{m,\lambda+1}^{i}$ after training round $\lambda$ (lines \ref{line:6}-\ref{line:9}). These parameters are then transmitted to their cluster's PS for aggregation. 

After $m$ rounds of training in each satellite cluster, each PS forwards its aggregated parameters to its associated ground station for global aggregation, producing the updated model parameters $w_{m+1}$ (lines \ref{line:12}-\ref{line:13}). During global aggregation, satellite clusters monitor whether the number of dropped-out satellites $C^d$ exceeds a predefined threshold, triggering re-clustering when necessary (lines \ref{line:14}-\ref{line:18}). Finally, ground station broadcasts the global parameters to all affiliated satellites in their clusters, completing the hierarchically clustered FL process (lines \ref{line:21}-\ref{line:23}).

To accelerate the convergence of the global model, we assign weights to clients based on the quality of their model updates. The quality is evaluated using the loss value of the client's local model. Let $L_i$ denote the loss value of the client $i$. The weight $p_i$ assigned to client $i$ is given by:
\begin{equation}
p_i = \frac{\frac{1}{L_i}}{\sum_{ i \in C^k} \frac{1}{L_i}}
\label{eq:p}
\end{equation}

\begin{comment}
The local updates from the satellite clients are then weighted and aggregated according to the weight parameter $p_i$ by:
\begin{equation}
w_{m+1} = w_m + \sum_{i \in U_k} p_i Q_m^i(w_{m+1}^{i} - w_m)
\end{equation}
\end{comment}
%If $p_i$ is 0, the client has not been selected by PS.

\subsection{Satellite-clustered parameter server selection algorithm}

We introduce a satellite-clustered parameter server selection algorithm that partitions the original satellite network topology into a predefined number of clusters $K$, optimizing the clustering process. Our algorithm iteratively refines the cluster centroids and the membership of associated satellites. Initially, $K$ centroids are randomly selected from the satellite location data. These locations are typically derived from geographic coordinates based on the satellite location parameters, i.e., inclination and orbital altitude. Each satellite is assigned to the nearest cluster centroid using the Euclidean distance metric, thereby forming initial clusters. The Euclidean distance between a satellite position vector
$\mathbf{C}^i = \{C_{1}^i, C_{2}^i, \ldots, C_{n}^i \}$ and a centroid $\mathbf{K}^j = \{K_{1}^j, K_{2}^j, \ldots, K_{n}^j \}$ is calculated as:
\begin{equation}
d(\mathbf{C}^i, \mathbf{K}^j) = \sqrt{\sum_{k=1}^{n} (C_{k}^i - K_{k}^j)^2}
\end{equation}

In the next update step, our algorithm recalculates the centroids by computing the mean position of all satellites assigned to each cluster. This process effectively repositions the centroids to more accurately represent the distribution of their associated satellites. For each cluster $K_k^i$, the new centroid $\mathbf{K}^j$ is obtained by:
\begin{equation}
\mathbf{K}^j = \frac{1}{|K^j|} \sum_{\mathbf{C}^i \in K^j} \mathbf{C}^i
\end{equation}
where $|K^j|$ represents the number of satellites in cluster $K^j$. The iterative process continues until the centroids stabilize, indicating their positions no longer change significantly between iterations. This indicates that the algorithm has converged to a local optimum. The convergence criterion is given by:
\begin{equation}
\sum_{j=1}^{|K|} \|\mathbf{K}_{\text{new}}^j - \mathbf{K}_{\text{old}}^j\|^2 < \epsilon
\end{equation}
where $|K|$ represents the number of clusters, and $\epsilon$ is a small positive number indicating stability in centroid positions.
The satellite nearest to the cluster centroid is designated as the PS for the respective cluster.

\subsection{Meta-learning-driven satellite re-clustering algorithm}

In dynamic satellite federated learning, the diverse training objectives of satellite clients, combined with their frequent network participation changes, can hinder model convergence and increase resource consumption. As a result, achieving acceptable performance requires substantial time and a large number of data samples.
%In dynamic satellite federated learning, the diverse training goals of satellite clients, coupled with their frequent network participation changes, can impede model convergence, resulting in heightened resource utilization. This necessitates a significant investment of time and data samples to attain acceptable performance. %the different training objectives of satellite clients, along with their frequent joining or leaving the network, can slow down model convergence, leading to increased resource consumption. This requires a large amount of time and data samples to achieve acceptable results. 

To address this challenge, we propose a satellite re-clustering algorithm based on meta-learning, extending the original satellite-clustered parameter server selection algorithm. When a new satellite joins the network, it inherits model updates from the head node of a specified cluster, rather than starting training from scratch. The core idea of the MAML approach is to identify a set of meta-initialization parameters that enable the model to achieve strong performance with just one or two gradient updates, even with a small number of new task examples.

First, we sample satellite clients from different clusters denoted as $ S = \{S_1, S_2, \dots, S_n\} $. Each satellite client is assigned a task $\textit{TK}_i$, which consists of a dataset $D_i$  and a loss function $L_{S_i}(w)$. The objective is to minimize the loss of the model on the task $\textit{TK}_i$. Then, an inner-loop adaptation is performed for each selected satellite node to fine-tune the global model $w$ by:
\begin{equation}
w_i' = w - \alpha \nabla_{w} L_{S_i}(w)
\end{equation}
where $\alpha$ is the local learning rate. Finally, an outer-loop meta-update is applied to aggregate the model updates from different satellite nodes, updating the global initialization by:
\begin{equation}
w^{t+1} = w^{t} - \beta \sum_{i \in S} \nabla_{w} L_{S_i}(w_i')
\end{equation}
where $\beta$ is the meta-learning rate, $w^{t}$ is the parameter of the current global model.

\section{Experimental Evaluation}~\label{sec:evaluation}

\subsection{Experimental setup}

We evaluate the performance of the proposed \algname using our established testbed, which simulates LEO satellite networks while considering the orbital movement of the satellites. The satellites are evenly distributed across each orbit, operating at an altitude of 1300 km with an inclination angle of $53^{\circ}$. The base station maintains a minimum elevation angle of $10^{\circ}$. Each client is trained using a small batch SGD with a batch size of 64, employing the LeNet model. The initial learning rate is set to 0.01, and clients train the FL model using the MNIST dataset and the CIFAR-10 dataset. The original dataset is partitioned into different subsets corresponding to the number of satellite clients. After dividing the satellite clients into $K$ clusters, local training is conducted for 500 rounds within each cluster. The communication and computation parameters in satellite networks follow those outlined in~\cite{ZhuJSAC23, ZhangIoT23}.
%The CPU cycle frequency of the satellite is $ f_i = 50 \, \text{GC/s}  (GC =  10^9 $ cycles)~\cite{ZhuJSAC23}, the transmission bandwidth is $ B_i = 27 \, \text{GHz} $ (Ka-band), the satellite transmission power is $ p_i = 30 \, \text{dBW} $~\cite{ZhangIoT23}, and the noise power density is $ N_0 = -174 \, \text{dBm/Hz} $~\cite{ZhuJSAC23}. 
We configure identical learning rates of $10^{-3}$ for both the inner ($\alpha$) and outer loops ($\beta$) in the satellite re-clustering algorithm. In terms of satellite mobility, we assume that satellites at the same latitude maintain their relative positions, and the ground station can connect at least one satellite cluster throughout the FL process.

\begin{figure}[tb!]
	\centering
	\begin{minipage}[b]{.493\columnwidth}
		\centering
		\includegraphics[width=\columnwidth]{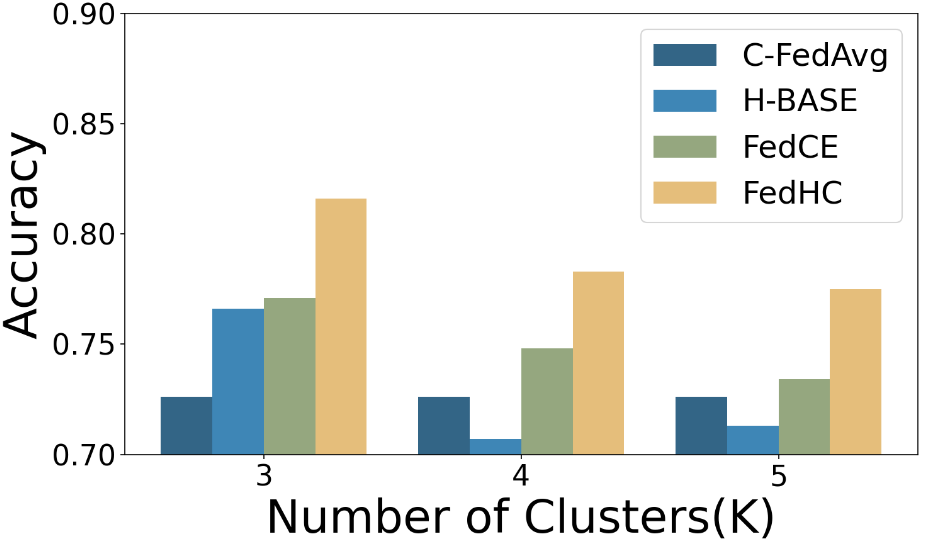}
		\subcaption{MNIST}\label{fig:1_acc}
	\end{minipage}
        \begin{minipage}[b]{.493\columnwidth}
		\centering
		
		\includegraphics[width=\columnwidth]{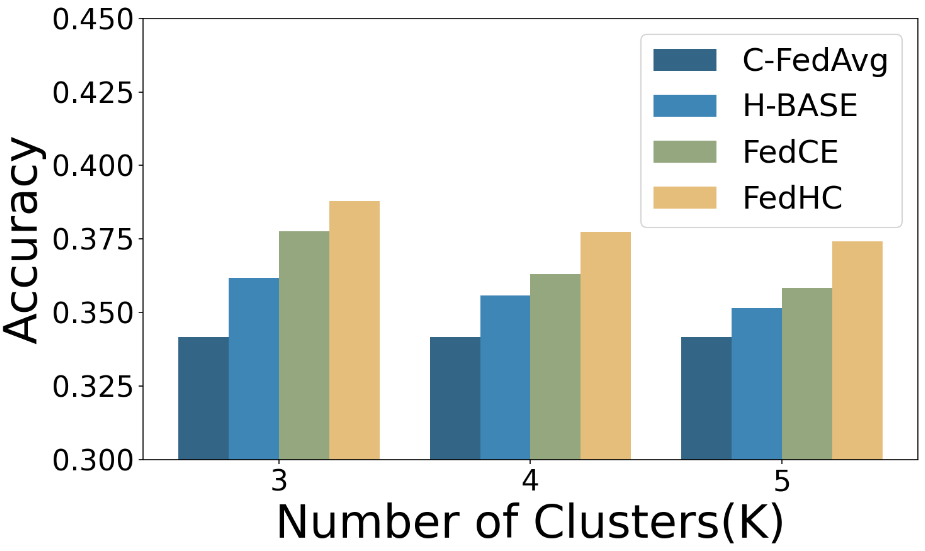}
		\subcaption{CIFAR-10}\label{fig:3_acc}
	\end{minipage}

	\caption{Accuracy performance with different datasets.}
    
	\label{fig:accuracy}
\end{figure}

%\subsection{Comparative methods}
We validate the effectiveness of our proposed \algname against the following three comparative methods.
\begin{itemize}

    %\item\textbf{FedAvg~\cite{McMahanPMLR17}}: is a traditional distributed learning algorithm
    
    %两个架构是否要引参考文献，但是这两个架构参考文献为一篇{Edge Selection and Clustering for Federated Learning in Optical Inter-LEO Satellite Constellation}
    
    \item\textbf{C-FedAvg}~\cite{ChenPIMRC23}: A centralized FL method based on FedAvg, where all data collected from each client is transmitted to a designated central satellite server for centralized learning.
    
    %\item\textbf{Distributed Learning Architecture~\cite{ChenPIMRC23}}: In distributed learning, LEO clients perform local training independently, without exchanging information through Inter-Satellite Links.

    \item\textbf{H-BASE}~\cite{LiuICC20}: A clustered FL method that randomly selects clients and performs training over a fixed number of intra-cluster aggregation iterations.
    
    \item\textbf{FedCE}~\cite{FedCE23}: A clustered FL method that partitions clients into clusters based on the distribution characteristics of their data.

\end{itemize}

%For the parameters in the comparison methods, we follow the optimal settings recommended in the original literature. 

%!!!排版参照图4，两两一排!!!

\subsection{Experimental results}

%横坐标为train round
%纵坐标为accuracy和loss
% k=3,4,5（6张图） 5个方法5条线  （折线图）
\textbf{Training convergence}: 
To validate the effectiveness of \algname, we fix the number of satellite clients at 800 and divide the satellite network into different clusters, i.e., $K$= 3, 4, 5. Then, we compare the performance of our method against other methods on the MNIST dataset after 300 rounds, and on the CIFAR-10 dataset after 1000 rounds. This evaluation aims to demonstrate the accuracy and convergence speed of \algname, highlighting its advantages over baseline methods. Figure~\ref{fig:accuracy} compares model accuracy across methods under various clustering configurations, evaluated over a fixed number of training rounds before full convergence. \algname consistently achieves higher accuracy than baseline methods within the same training round, particularly outperforming methods that do not incorporate meta-learning under satellite re-clustering conditions. 
%\figurename~\ref{fig:accuracy} shows the model accuracy of various methods under different clustering conditions, chosen based on a fixed number of rounds before complete convergence. As observed, \algname outperforms other baseline methods in terms of accuracy with the same number of rounds, while also demonstrating an improvement in convergence speed, especially compared to methods that do not incorporate meta-learning under satellite re-clustering conditions. 
In the following experiments, we present the results of all methods upon reaching the specified converged target accuracy thresholds  (MNIST: 80\%, CIFAR-10: 40\%), following the methodology outlined in~\cite{TMC23}.

\begin{table}[tb!]
\centering
\caption{Performance comparison of different methods.}
\label{tab:results}
\begin{threeparttable}
\begin{tabular}{l|ccl|ccl|ccl}
\hline
\multicolumn{1}{c|}{\multirow{2}{*}{\begin{tabular}[c]{@{}c@{}}Method\\ (\textbf{MNIST})\end{tabular}}} & \multicolumn{3}{c|}{\textit{K=3}}                    & \multicolumn{3}{c|}{\textit{K=4}}                   & \multicolumn{3}{c}{\textit{K=5}}                  \\ \cline{2-10} 
\multicolumn{1}{c|}{}                                                                             & Time                       & \multicolumn{2}{c|}{Energy}  & Time                      & \multicolumn{2}{c|}{Energy}  & Time                     & \multicolumn{2}{c}{Energy}  \\ \hline
C-FedAvg~\tnote{1}                                                                                          &     8184                     & \multicolumn{2}{c|}{3697}     &  8184                       & \multicolumn{2}{c|}{3697}     &         8184               &  \multicolumn{2}{c}{3697}     \\
H-BASE                                                                                            & 4734                     & \multicolumn{2}{c|}{2301} & 5687                    & \multicolumn{2}{c|}{2633} & 5055                   & \multicolumn{2}{c}{2761} \\
FedCE                                                                                            & 3945                     & \multicolumn{2}{c|}{1889} & 3696                        & \multicolumn{2}{c|}{1923}     &  2906                      & \multicolumn{2}{c}{2160}     \\
FedHC                                                                                             & \textbf{3124}                        & \multicolumn{2}{c|}{\textbf{1720}}     & \textbf{3554}                        & \multicolumn{2}{c|}{\textbf{1863}}     & \textbf{2527}                       & \multicolumn{2}{c}{\textbf{1889}}     \\ \hline\hline
\multicolumn{1}{c|}{\multirow{2}{*}{\begin{tabular}[c]{@{}c@{}}Method\\ (\textbf{CIFAR-10})\end{tabular}}} & \multicolumn{3}{c|}{\textit{K=3}}                    & \multicolumn{3}{c|}{\textit{K=4}}                   & \multicolumn{3}{c}{\textit{K=5}}                  \\ \cline{2-10} 
\multicolumn{1}{c|}{}                                                                             & \multicolumn{1}{c}{Time}   & \multicolumn{2}{c|}{Energy}  & Time                      & \multicolumn{2}{c|}{Energy}  & Time                     & \multicolumn{2}{c}{Energy}  \\ \hline
C-FedAvg~\tnote{1}                                                                                          & \multicolumn{1}{c}{13477}     & \multicolumn{2}{c|}{6637}     & \multicolumn{1}{c}{13477}    & \multicolumn{2}{c|}{6637}     & \multicolumn{1}{c}{13477}   & \multicolumn{2}{c}{6637}     \\
H-BASE                                                                                            & \multicolumn{1}{c}{9346}     & \multicolumn{2}{c|}{6841}     & \multicolumn{1}{c}{10129}    & \multicolumn{2}{c|}{5381}     & \multicolumn{1}{c}{12003}   & \multicolumn{2}{c}{5016}     \\
FedCE                                                                                            & \multicolumn{1}{c}{8315}     & \multicolumn{2}{c|}{5397}     & \multicolumn{1}{c}{7931}    & \multicolumn{2}{c|}{4674}     & \multicolumn{1}{c}{7258}   & \multicolumn{2}{c}{4451}     \\
FedHC                                                                                             & \multicolumn{1}{c}{\textbf{7873}}     & \multicolumn{2}{c|}{\textbf{5013}}     & \multicolumn{1}{c}{\textbf{7509}}    & \multicolumn{2}{c|}{\textbf{4259}}     & \multicolumn{1}{c}{\textbf{6942}}   & \multicolumn{2}{c}{\textbf{4148}}     \\ \hline
\end{tabular}
   \begin{tablenotes}
       \footnotesize
       %\item[1] PT: Processing Time (secs), TEC: Total Energy Cost (J).
       \item[1] As a centralized method, C-FedAvg achieves uniform performance across all scenarios since it operates without distributed collaboration ($K$=1). 
       
       %C-FedAvg records a same performance for all cases since it is a centralized method, i.e., $K$ = 1.
     \end{tablenotes}
     \end{threeparttable}
  
\end{table}

\textbf{FL processing time and energy consumption:} Table~\ref{tab:results} presents the total processing time (seconds, by Equation~\ref{eq:time}) and total energy consumption (J, by Equation~\ref{eq:energy}) across various clustering configurations and datasets. The results demonstrate that \algname outperforms all baselines on both datasets. In terms of processing efficiency, our method significantly reduces overall processing time compared to the three baseline methods. For example, with $K$=5 on the MNIST dataset, \algname demonstrates a twofold reduction in processing time against C-FedAvg. This improvement stems from hierarchical cluster FL, which deploys multiple parameter servers to enable parallelized model training across clusters, thereby drastically reducing communication time. This is because hierarchical cluster FL deploys multiple parameter servers, enabling parallelized model training across clusters and drastically reducing communication time. Compared to H-BASE and FedCE, which rely on static clustering or slower convergence strategies, \algname reduces processing time by up to 40\%. This improvement is driven by the meta-learning-driven satellite re-clustering algorithm, which accelerates FL convergence.

Meanwhile, for $K$=5 in the MNIST dataset, our method reduces total energy consumption by over twofold compared to C-FedAvg. In other cases, \algname consistently consumes approximately 1.5 times less energy than C-FedAvg. These results demonstrate the superior energy efficiency of our hierarchical clustered FL framework over centralized C-FedAvg. Furthermore, our method exhibits a significant advantage over other federated clustering methods, primarily due to two key innovations. First, the satellite-clustered parameter server selection algorithm optimizes transmission energy by strategically choosing parameter servers with favorable geographical positions and communication capabilities. Second, the meta-learning-driven satellite re-clustering algorithm accelerates model convergence, drastically reducing energy consumption during training. These features make our framework particularly well-suited for resource-constrained satellite networks, where efficient communication and rapid convergence are critical for minimizing energy consumption.

\section{Conclusion}~\label{sec:conclusion}
In this paper, we propose a hierarchical clustered FL framework \algname that optimizes both processing time and energy efficiency for satellite networks. To address the dynamic topology caused by high-speed LEO satellite mobility and the unstable ground-to-satellite communication links, we introduce a satellite-clustered PS selection algorithm that optimizes network partitioning and strategically selects in-orbit PS to reduce transmission latency, and a meta-learning-driven satellite re-clustering algorithm that dynamically adapts cluster configurations during aggregation to accelerate global model convergence. Experimental results demonstrate that \algname achieves significant reductions in the processing time and energy consumption compared to the comparative FL-based methods while ensuring the accuracy of the model. Our future work aims to enhance data integrity and confidentiality during collaborative learning by integrating advanced privacy-preserving mechanisms such as differential privacy to further solidify \algname’s applicability in security-critical applications.

%In this paper, we propose a hierarchical clustered federated learning framework \algname for satellite networks. This framework aims at minimizing FL task completion time and energy cost while maintaining model accuracy. To address the challenge posed by the high-speed movement of LEO satellites that prevents continuous communication with ground PS, we design a satellite-clustered parameter server selection algorithm to partition the satellite network and select a suitable PS  to accelerate the FL process.And use the satellite re-clustering algorithm during satellite aggregation to accelerate the convergence of the model. To validate the effectiveness of our proposed \algname, we conducted extensive experiments simulating the LEO satellite network using real-world data. The obtained results confirm that, compared to the existing FL frameworks, \algname can significantly reduce task processing time and energy cost while ensuring the accuracy of the model. Our future work aims to integrate advanced privacy protection technologies to protect the integrity and confidentiality of data during the learning process. 

\begin{comment}
\section*{Acknowledgment}

The preferred spelling of the word ``acknowledgment'' in America is without 
an ``e'' after the ``g''. Avoid the stilted expression ``one of us (R. B. 
G.) thanks $\ldots$''. Instead, try ``R. B. G. thanks$\ldots$''. Put sponsor 
acknowledgments in the unnumbered footnote on the first page.
\end{comment}

\bibliographystyle{ieeetr} 
\bibliography{ref}

\begin{comment}
%表格暂时写在上面，方便对照
\begin{table}[ht]
\centering
\caption{Notations and Descriptions\note{[REMOVE THIS TABLE AT LAST]}}
\begin{tabular}{|c|l|}
\hline
\textbf{Inputs} & \textbf{Description} \\ \hline
$C$ & the set of candidate clients in FL \\ \hline
$K$ & the set of clusters \\ \hline
$D_i$ & Each client $c_i$ holds a local dataset size \\ \hline
$p_k$ & the cluster head of cluster satellite $k$ \\ \hline
$S_k$ & the clients belongs to cluster $k$ \\ \hline
$G$ & ground station number $G$ \\ \hline
$g_{i}^{p_k}$ & the cluster$p_k$ belongs to ground station i  \\ \hline
$t^{cmp}_i$ & computation time of client $i$ for one intra-cluster aggregation \\ \hline
$t^{com}_i$ & communication time of client $i$ for one intra-cluster aggregation \\ \hline
$t^{com}_k$ & communication time of cluster head $l_k$ for one inter-cluster aggregation \\ \hline
$n_k$ & Maximum selected number of clients per cluster per global round \\ \hline
\textbf{Outputs} & \textbf{Description} \\ \hline
$x^k_i$ & whether client $i$ belongs to cluster $k$ \\ \hline
$y^t_i$ & whether or not select client $i$ at global round $t$ \\ \hline
$\tau_{i}^{k}$ & local iterations given to client ci in the k-th communication round \\ \hline
\end{tabular}
\end{table}
\end{comment}

\end{document}